# All-optical wavelength multicasting in quadruple resonance-split coupled Silicon microring cavity

AWANISH PANDEY[1],[*] AND SHANKAR KUMAR SELVARAJA[1]

[1]*Centre for Nano Science and Engineering, Indian Institute of Science, Bengaluru 560012, India*

[*]*Corresponding author: awanish@iisc.ac.in*

**We demonstrate an all-optical four-channel wavelength multicasting in a coupled Silicon microring resonator system. The scheme is based on two-photon absorption induced free carrier dispersion in Silicon. The coupled cavity facilitates resonance splitting that is utilized as individual channels for multicasting. Using the split resonances, we achieve an aggregate multicasted data rate of 48 Gbps (4×12 Gbps). Moreover, we also present a detailed analysis and performance of the multicasting architecture.**

Wavelength multicasting is an essential functionality in optical networks that facilitates efficient and simultaneous routing of information from a single source to desired multiple recipients [1–3]. It involves replicating an input optical signal onto many different wavelengths and hence enabling data to be simultaneously forwarded to more than one recipient. Moreover, multicasting can also be employed to accomplish several other signal processing appliations such as radio frequency reciever [4] and tapped delay lines [5].

Silicon photonics based on-chip optical methods have emerged as a promising candidate for multicasting. They offer most compact and economical solution combined with multitudes of functionality compared to other competing platforms like wavelength selective switches and micro-opto-electromechanical systems (MOEMS) [6]. Despite weak non-linear properties of Si; low propagation loss, high light confinement and power density in Si waveguides allows enhanced non-linear interactions even at relatively low power levels that can be exploited for various non-linear optical applications[7]. Wavelength multicasting has been demonstrated using Si wire waveguides utilising different non-linear processes like cross-phase modulation [8], four wave mixing in Si, SOA and fiber [9–12] and also in different hybrid Si platforms like Indium-Phosphideon-Si waveguides [13].

In all the above demonstrations, Si platfrom uses dispersion engineered long Si waveguides. The effect of non-linearity in Si waveguides can be enhanced by using resonant structures like microring resonators (MRRs). The intensity build-up in MRRs lowers the power requirement even more so that non-linear effects can be achieved for optical power in the waveguide as low as 3mW [14]. Multicasting has also been demonstrated using MRRs where the resonance shift is achieved through Free Carrier Dispersion (FCD) [15, 16]. Free carriers are generated at MRR resonance wavelength as a consequence of two−photon absorption (TPA) in Silicon. This leads to Free Carrier Dispersion (FCD) that can then be utilised to attain multicasting via MRR resonances [16].

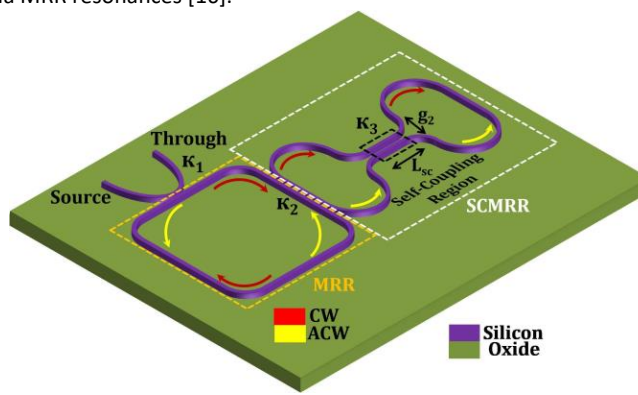

**Fig. 1.** Schematic of a self-coupled microring resonator.

Since MRR resonances act as multicasting channels, it becomes necessary to have small Free Spectral Range (FSR) to enhance the number of channels. However, the inverse relationship between cavity size and FSR mandates MRR to have large geometrical length. Such designs not only compromises the compactness of cavity but also renders it susceptible towards propagation loss.

We have earlier shown that splitting the resonances is a potential solution where a cavity resonance could be selectively splitted into closely spaced reconences [17]. Such splitting of resonances have already been demonstrated and utilized in areas like photonic links [18]. In this paper, we propose a coupled cavity system that can split a cavity resonance into four closely spaced sub-resonances. We demonstrate four channel wavelength multicasting by transmitting data at 12 Gbps per channel and an aggregate data rate of 48 Gbps. We report comprehensive simulation and experimental analysis of the proposed structure and analyse the quality of multicasting channels using eye diagram and bit error rate analysis.

A schematic of the proposed coupled cavity system is shown in Fig. 1. It is an assembly of a conventional MRR externally loaded with a Self−Coupled MRR (SCMRR). A detailed discussion on the design and working principle of the SCMRR is presented in ref[17]. A typical MRR has mode propagating in either clockwise or anti-clockwise direction and resonates at wavelengths that phase match with the cavity. However, SCMRR has the ability to excite both clockwise and anti-clockwise propagating modes simultaneously due to the addition of a Self-Coupling Region (SCR) in the structure (Fig. 1). Interference between these two modes results in a split of the resonances. The SCMRR self-coupling co-efficient $\kappa_3$ can control the spacing between the resonances. A simple MRR coupled to another MRR can also split the resonance when phase matched [19]. The coupled MRR introduces optical path length difference between the single MRR mode and combined cavity hybrid mode and depending on the path length difference; selective resonances can be split. In our design, we achieve both types of splitting simultaneously, resulting in four cavity resonances by coupling a conventional MRR to a SCMRR.

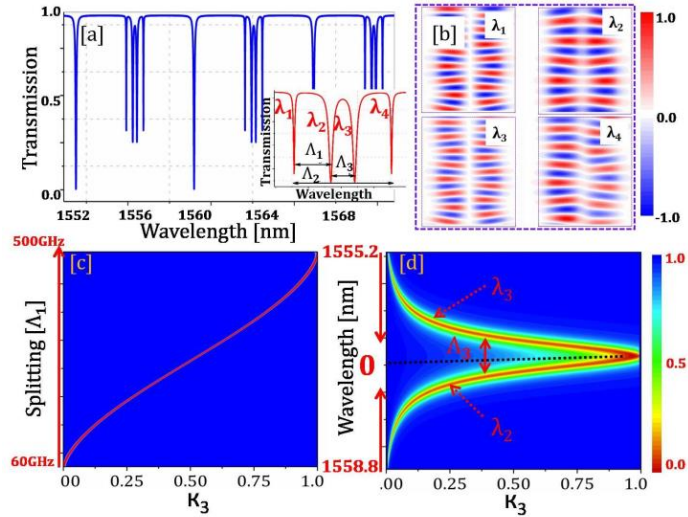

**Fig. 2.** Simulation results summary; (a) transmission spectra of the proposed cavity, (b) Field distribution in the self-coupling region at four split wavelengths, (c) and (d) variation of $\Lambda_1$ and $\Lambda_3$ respectively with $\kappa_3$

Simulated results of the proposed coupled cavity are shown in Fig. 2. The circumference of the SCMRR (87 $\mu$m) was chosen to be half of the MRR (174 $\mu$m). A coupling length of 10$\mu$m between MRR and SCMRR is chosen to allows phase matching of alternate resonances of MRR with SCMRR. As shown in Fig. 2a, alternate resonances have undergone quadruple splitting at wavelengths $\lambda_1$, $\lambda_2$, $\lambda_3$ and $\lambda_4$ whereas other resonances are unperturbed. Here, we define FSR as the distance between resonances of MRR when it not coupled to SCMRR ($\kappa_2 = 0$). The splitting is characterized by calculating $\Lambda_1$, $\Lambda_2$ and $\Lambda_3$ as shown in the inset of Fig. 2a. Fig. 2b shows a 2D-FDTD simulation of the field distribution in the coupling region. The fields are observed to have an anti-symmetric coupling at $\lambda_1$ and $\lambda_3$ whereas symmetric coupling at $\lambda_2$ and $\lambda_4$. The conjugate symmetric and anti-symmetric field distribution at successive split wavelengths confirms that the splitting arises due to interference between two contra-propagating degenerate resonant modes of the cavity [20].

As mentioned earlier, self-coupling in SCMRR plays a crucial role in resonance splitting. Fig. 2c and 2d show the effect of $\kappa_3$ on $\Lambda_1$ and $\Lambda_3$. For $\kappa_3 = 0$, $\Lambda_1$ is entirely from co-propagating SCMRR mode interference (60 GHz). As the power in anticlockwise mode is increased by increasing $\kappa_3$ the splitting starts to increase as shown in Fig. 2c and becomes equal to 500 GHz at $\kappa_3 = 1$, which is the FSR of the MRR. The $\Lambda_3$, however, shows a decreasing trend with $\kappa_3$ and reaches 0 for $\kappa_3 = 1$ as shown in Fig. 2d. It is worth noting that at $\kappa_3 = 1$, SCMRR imparts an additional $\pi$ phase shift at the SCR and as a result the entire spectrum of the coupled cavity shifts by FSR/2.

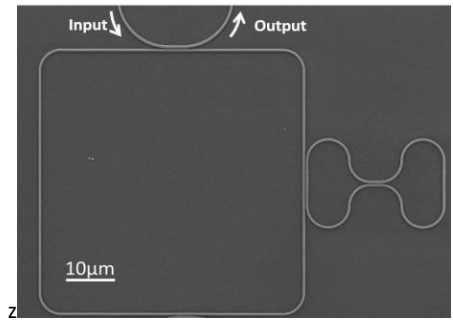

**Fig. 3.** SEM image of the fabricated device.

The proposed coupled cavity structure was patterned on a Silicon-On-Insulator wafer with 220 nm Si layer and 2 $\mu m$ thick buried oxide using electron-beam lithography, and dry etch process. The pattern in the resist is transferred into 220 nm Si layer using a Fluorine based chemistry using an inductive coupled plasma dry etch process. SEM image of fabricated device is shown in Fig. 4. Grating fiber-chip couplers were fabricated to couple the light in and out of the device. A 2 $\mu m$ of $SiO_2$ was deposited using plasma enhanced chemical vapour deposition process to isolate the device from the metal microheaters. The micro-heaters were processed using *Au* lift-off process to tune the self-coupling of the SCMRR.

The spectral response of the fabricated device is shown in Fig. 3a. The measurements were performed using a tunable laser and a power meter. The device shows an alternate pattern of split resonance as discussed earlier. Fig. 3b shows the unperturbed and splitting response of the fabricated device. However, at higher wavelengths we observe splitting of the unperturbed as well (Fig. 3c). We attribute this splitting to reflection at the MRR-SCMRR coupling region. Since the coupling length between MRR and SCMRR is 10 $\mu m$, the linewidth variation of the waveguides increases the back-reflections at higher wavelengths. The mode size becomes larger that could excites a counter-propagating mode from the MRR-SCMRR coupling region. Such back reflec tions can split the unperturbed MRR resonances as highlighted in Fig. 3(c) where the resonance not phase matched with SCMRR has split.

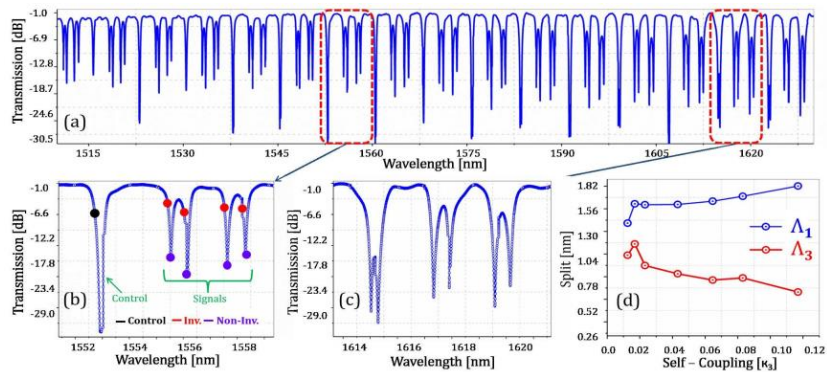

**Fig. 4.** (a) Spectral transmission curve of coupled cavity system (b) placement of control and signal wavelengths for multicasting (c) splitting of unperturbed resonances due to non-ideal coupling between MRR and SCMRR and (d) variation of splitting with different $\kappa_3$.

As discussed previously, the spacing between the split resonances can be varied by tuning the $\kappa_3$ of the SCMRR. Fig. 3d depicts the evolution of the splitting; $\Lambda_1$ and $\Lambda_3$, with $\kappa_3$. We observe an increase in $\Lambda_1$ from 180 GHz to 220 GHz while $\Lambda_3$ decreases from 165 GHz to 70 GHz for $\kappa_3$ variation of 12%, which matched well with our simulation (Fig.2). When the signal wavelength is parked at the split resonances and the modulated control (with NRZ random bits) at the slope of the unperturbed resonance, a non-inverted multicasting is obtained ( Fig. 3b purple dots). If the control carries a bit '1', free carriers are generated in Si that decreases the effective index. The FCD induced decrease in effective index blue shifts the resonance which in turn increases the transmission of the signal. In this way bit '1' is passed as a bit '1' and since bit '0' will not generate any free carrier, the signal will remain aligned with resonance and hence passed as bit '0'. However, when the signals are aligned at the slope of the split resonances as shown in Fig. 3b (red dots), the scheme would generate an inverted signal. Bit '1' induces blue shift and aligns the signal to the resonance dip whereas bit '0' keeps the alignment of signal on the slope. It results in inversion of the bit pattern, and an inverted multicasting is achieved.

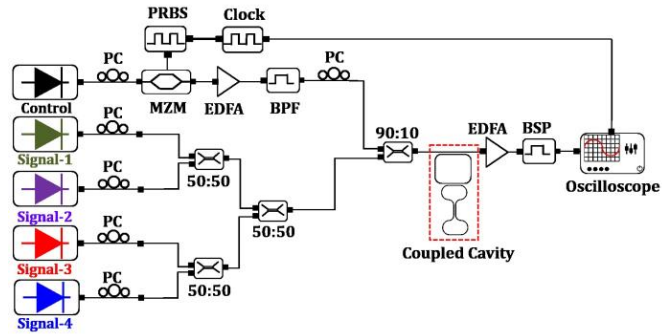

**Fig. 5.** Experimental setup for wavelength multicasting. MZM: MachZehnder Modulator, PRBS: Pattered Random Bit Sequence. EDFA: Erbirum Doped fiber Amplifier, BPF: Band Pass Filter, BSP: Band Stop Filter, PC: Polarization Controller.

The experimental set-up for wavelength multicasting is shown in Fig. 5. The unperturbed and split resonance is shown in detail in Fig. 3b is used to demonstrate multicasting. The placement of the control and signal wavelength is labelled in Fig. 3b. The extinction and Q-factor of the control and four signals are (27.8 dB, 7,000), (10.95 dB, 11,300), (14.11 dB, 9,100), (13.27 dB, 9,300) and (10.9 dB, 11,700) respectively. A Control is modulated with $2^{15} - 1$ pseudo–random bit stream and amplified using an EDFA followed by a bandpass filter (BPF). The control optical power is tuned by changing the EDFA current and BPF is used to suppress the spontaneous emission noise from the EDFA. Once the minimum power required to achieve multicasting is determined, the EDFA current is fixed to maintain the signal-to-noise ratio from the EDFA throughout the experiment. The control is then combined with the four signal wavelength using a 90:10 coupler and finally coupled to the device using a grating coupler. A control power of 0 dBm and a signal power of -12 dBm is used just before the grating couplers. At the output, the control and signals were amplified again using an EDFA and the control is filtered using a bandstop filter. The filtered signal is analysed using a digital oscilloscope. A variable optical attenuator is used after the BPF for BER measurements.

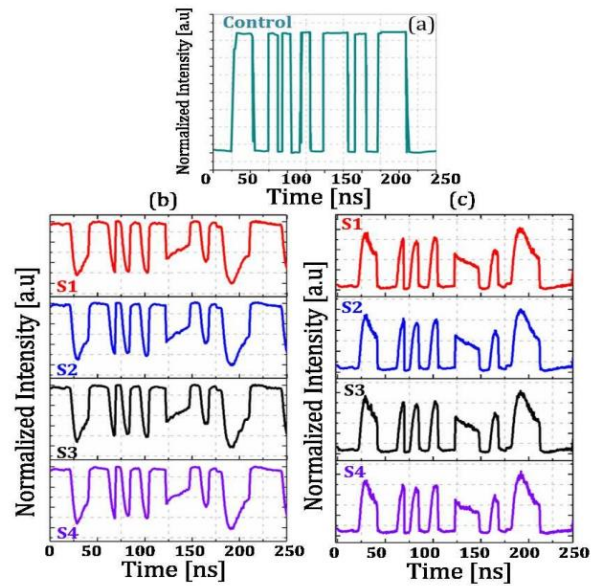

**Fig. 6.** (a) Control bit pattern and four channel multicasting for (a)inverted and (b) non-inverted signal. (S: Signal)

Fig. 6 shows the multicast output for the four signals for both, inverted and non-inverted cases, at 1.8 Gbps. It has been noted from the multicast output that the shape distortion of the isolated '1' is severe than consecutive bit '1'. The distortion can be attributed to the free carrier lifetime and the bit rate. Since isolated '1's have shorter extinction duration compared to successive '1's, the extinction of isolated '1's is observed to be lower than the successive '1's. Furthermore, with successive '1's we observe distortion in the pulse shape due to higher recombination then generation rate. The distortion of isolated '1's will be even more severely affected when the frequency of operation is much larger than the carrier generation frequency, which is observed in high data rate and signal quality characterisation presented in the following section.

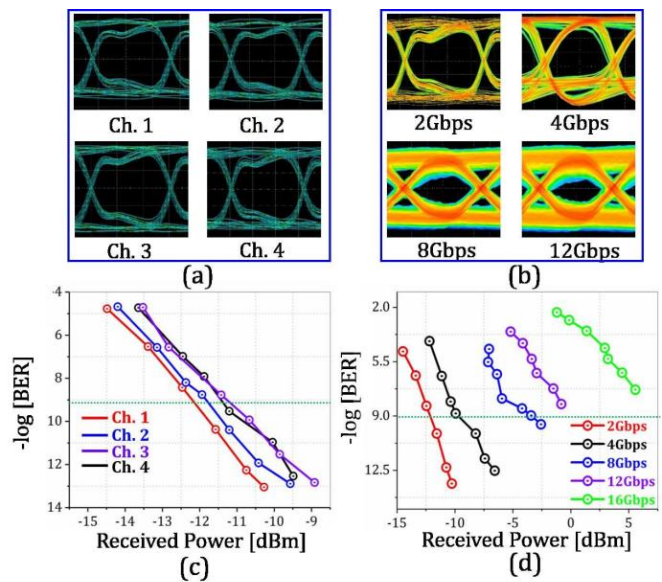

**Fig. 7.** (a) and (c) procide eye diagram and bit error rate measurements at 1.8Gbps whereas measurements for higher frequencies are procided in (b) and (d).

Fig. 7a shows the eye diagram for the four channels operating at 1.8 Gbps and Fig. 7c shows the corresponding BER for the channels. The eye remains wide open for all the channels. The BER also remains almost same with a power variation of around
1 dB for error-free operation, and this is attributed to extinction difference between the channels. We take standard error-free transmission as 1-bit error in $10^9$ bits. With increasing data rate we could observer degradation in the eye beyond 12 Gbps (Fig. 7b). The same is observed in BER power penalty plot in Fig. 7d. BER degrades progressively with data rate. However, even at 12 Gbps, the BER is above the Forward Error Correction (FEC) limit. With this demonstration, we present for the first time an FCA assisted multicasting in Si operating at an aggregated data rate of 48 Gbps (4×12 Gbps).

In conclusion, we have experimentally demonstrated wavelength multicasting using a coupled Si coupled microring resonator. A quadruple resonance splitting is achieved in a coupled MRR system that is closely spaced with high-quality factor. We have shown tunable nature of the splitting as well. The split resonance along with the unperturbed resonance is utilised to demonstrate wavelength multicasting by exploring TPA induced free carrier dispersion in the cavity. We have demonstrated an aggregate data rate of 48 Gbps (4×12 Gbps). A detailed signal analysis is also presented. The speed of operation could be further increased by tailoring the carrier lifetime in the Si waveguide.

The authors thank Ministry of Electronics and Information Technology, Defence Research and Development Organisation, Government of Indian for the support.